\documentclass[a4paper]{jpconf}
\usepackage{hyperref}
\usepackage{amsmath}
\usepackage[braket,qm]{qcircuit}
\usepackage{xcolor}
\usepackage{lineno}
\usepackage{placeins}
\usepackage{graphicx}
\usepackage{xcolor}
\hypersetup{
    colorlinks,
    linkcolor={red!50!black},
    citecolor={blue!50!black},
    urlcolor={blue!80!black}
}

\begin{document}
\title{Quantum algorithms for charged particle track reconstruction in the LUXE experiment}

\author{Arianna Crippa$^{1,2}$, Lena Funcke$^{3,4}$, Tobias Hartung$^{5}$, Beate Heinemann$^{6,7}$, Karl Jansen$^{1,8}$, Annabel Kropf$^{6,7}$, Stefan K\"uhn$^{1,8}$, Federico Meloni$^6$, David Spataro$^{6,7}$, Cenk T\"uys\"uz$^{1,2}$  and Yee Chinn Yap$^6$}

\address{$^1$ Deutsches Elektronen-Synchrotron DESY, Platanenallee 6, 15738 Zeuthen, Germany}
\address{$^2$ Institut für Physik, Humboldt-Universit\"at zu Berlin, Newtonstr. 15, 12489 Berlin, Germany}
\address{$^3$ Transdisciplinary Research Area ``Building Blocks of Matter and Fundamental Interactions'' (TRA Matter) and Helmholtz Institute for Radiation and Nuclear Physics (HISKP), University of Bonn, Nußallee 14-16, 53115 Bonn, Germany}
\address{$^4$ Center for Theoretical Physics, Co-Design Center for Quantum Advantage, and NSF AI Institute for Artificial Intelligence and Fundamental Interactions, Massachusetts Institute of Technology, 77 Massachusetts Avenue, Cambridge, MA 02139, USA}
\address{$^5$ Northeastern University - London, Devon House, St Katharine Docks, London, E1W 1LP, United Kingdom}
\address{$^6$ Deutsches Elektronen-Synchrotron DESY, Notkestr. 85, 22607 Hamburg, Germany}
\address{$^7$ Physikalisches Institut, Albert-Ludwigs-Universit\"at Freiburg, Hermann-Herder-Str. 3a, 79104 Freiburg, Germany}
\address{$^8$ Computation-Based Science and Technology Research Center, The Cyprus Institute, 20 Kavafi Street, 2121 Nicosia, Cyprus}

\ead{federico.meloni@desy.de\\ \textnormal{Preprint numbers:}
  DESY-23-045 , MIT-CTP/5481}

\begin{abstract}
  The LUXE experiment is a new experiment in planning in Hamburg, which will study Quantum Electrodynamics at the strong-field frontier. LUXE intends to measure the positron production rate in this unprecedented regime by using, among others, a silicon tracking detector. The large number of expected positrons traversing the sensitive detector layers results in an extremely challenging combinatorial problem, which can become computationally expensive for classical computers. This paper investigates the potential future use of gate-based quantum computers for pattern recognition in track reconstruction. Approaches based on a quadratic unconstrained binary optimisation and a quantum graph neural network are investigated in classical simulations of quantum devices and compared with a classical track reconstruction algorithm. In addition, a proof-of-principle study is performed using quantum hardware.
\end{abstract}


\section{Introduction}

The Laser Und XFEL Experiment (LUXE)~\cite{Abramowicz:2021zja} at DESY and the European XFEL (Eu.XFEL) aims at studying strong-field Quantum Electrodynamics (QED) processes in the interactions of a high-intensity optical laser and the 16.5~GeV electron beam of the Eu.XFEL ($e^{-}$-laser collisions), as well as with high-energy secondary photons. A strong background field is provided by a Terawatt-scale laser pulse and enhanced by the Lorentz boost of the electrons, allowing LUXE to explore a previously uncharted intensity regime.

\begin{figure}[htb]
\centering
\includegraphics[width=0.7\textwidth]{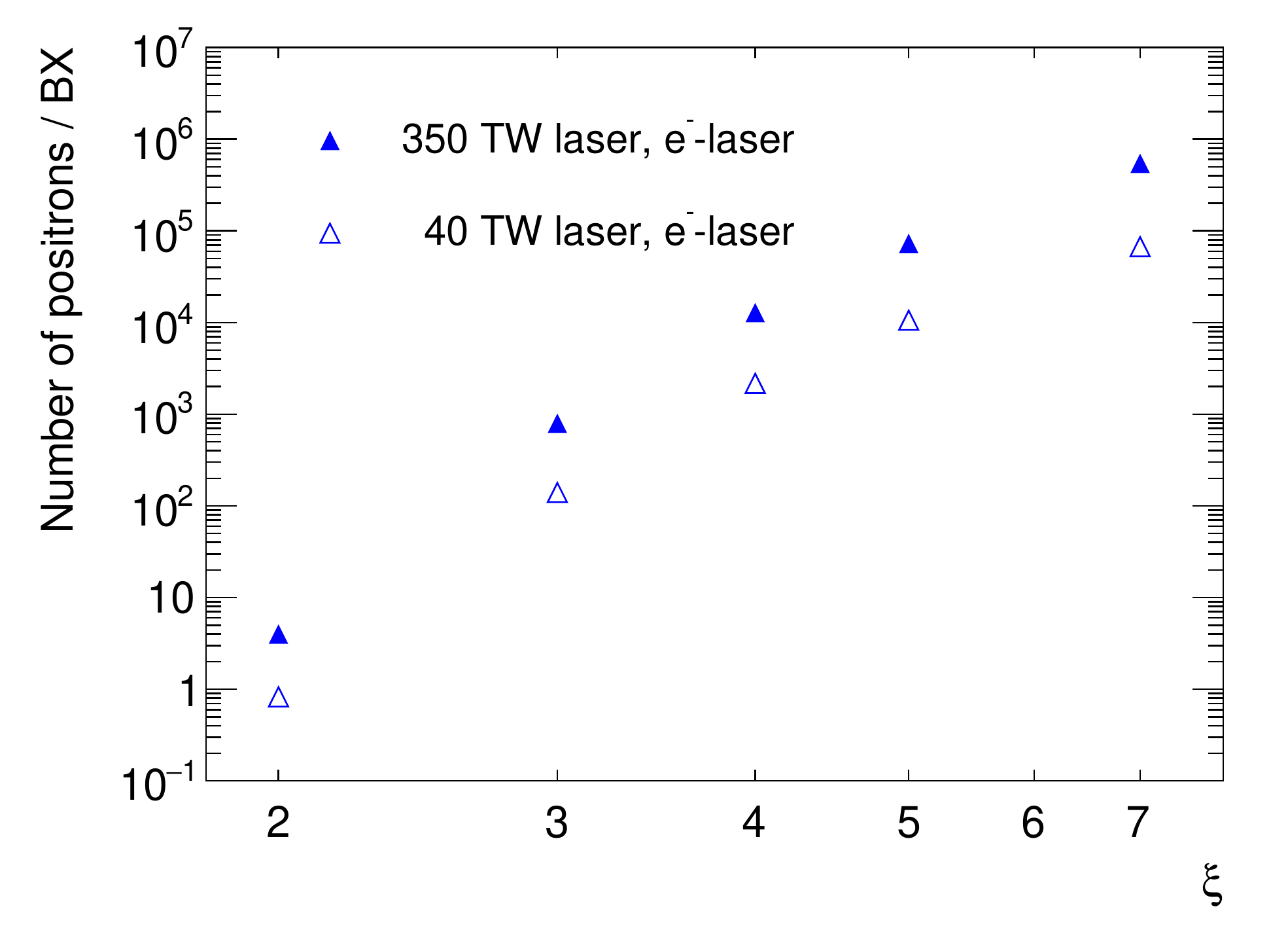}
\caption{Number of positrons per bunch crossing produced in $e^{-}$-laser collisions as a function of the laser field intensity parameter $\xi$, for different values of the laser power. Based on Ref.~\cite{Abramowicz:2021zja}, with additional simulated events.}
\label{fig:positronrate}
\end{figure}

In this regime, one of the main goals of the LUXE experiment is to measure the positron rate as a function of the laser intensity parameter $\xi$, defined as
\begin{equation}
  \xi=\sqrt{4\pi\alpha}\,\,\frac{\epsilon_L}{\omega_L m_e}=\frac{m_e \epsilon_L}{\omega_L \epsilon_{cr}},
  \label{eq:xi}
\end{equation}
where $\alpha$ is the fine structure constant, $\epsilon_L$ is the laser field strength, $\omega_L$ is the frequency of the laser, $m_e$ is the electron mass, and $\epsilon_{cr}=1.32 \times 10^{18}$~V/m is the critical field strength, also known as the Schwinger limit~\cite{Schwinger:1951nm}.
The measured positron rate will be compared to theoretical predictions. When considering electron-laser collisions, the dominant process is the non-linear Compton scattering~\cite{Nikishov:1964zza,Brown:1964zzb}. In non-linear Compton scattering, the incident electron absorbs multiple laser photons, emitting a Compton photon, which can then interact again with the laser field to produce electron-positron pairs~\cite{Breit:1934zz,Reiss:1962hr,narozhny69}. The expected number of positrons per bunch crossing (BX) as a function of $\xi$ spans over five orders of magnitude in the range shown in Figure~\ref{fig:positronrate}.

The measurement of the positron rate will be performed by a dedicated set of detectors comprising a silicon pixel tracker and a calorimeter. The wide range of expected positron rates poses a significant challenge to event reconstruction, especially within the tracker, where the large number of energy deposits could lead to finding spurious tracks that do not correspond to a real particle. The most relevant tracking challenge for this work is to maintain a linear dependence of the number of reconstructed tracks as a function of the number of charged particles in the event up to very high particle multiplicities. 

This work investigates the potential future use of gate-based quantum computers for pattern recognition in track reconstruction and compares the obtained performance to classical methods. 
Analogous studies have focused on track reconstruction in the proton-proton collision environments of the Large Hadron Collider and its upgrades, by using quantum annealers~\cite{Bapst:2019llh,Schwagerl:2023elf}, quantum associative memories~\cite{Shapoval:2019txi} or quantum graph neural networks~\cite{Tuysuz:2021oai}. A review of various quantum computing algorithms studied for charged particle tracking can be found in Ref.~\cite{Gray:2021bkw}. In this work, we present an update of our previous study of track reconstruction with quantum algorithms at LUXE~\cite{Funcke:2022dws,Crippa:2022hbi}.

This paper is organised as follows. A brief characterisation of the current proposed detector layout and the data-taking environment are given in Section~\ref{sec:detector}. The data sets used in this study are presented in Section~\ref{sec:datasets}, together with the dedicated simulation software. 
Section~\ref{sec:methods} presents the methodology used for the reconstruction of the simulated data. 
The results are discussed in Section~\ref{sec:results}, focusing first on classical simulations of quantum hardware and then presenting a set of studies performed on quantum hardware (ibm\_nairobi). The summary and conclusion are given in Section~\ref{sec:conclusions}, while an outlook on future developments and work is discussed in Section~\ref{sec:outlook}.

\section{The LUXE experiment}
\label{sec:detector}

\begin{figure}[h]
\centering
\includegraphics[trim={0 0 1cm 0.1cm},clip,width=0.6\textwidth]{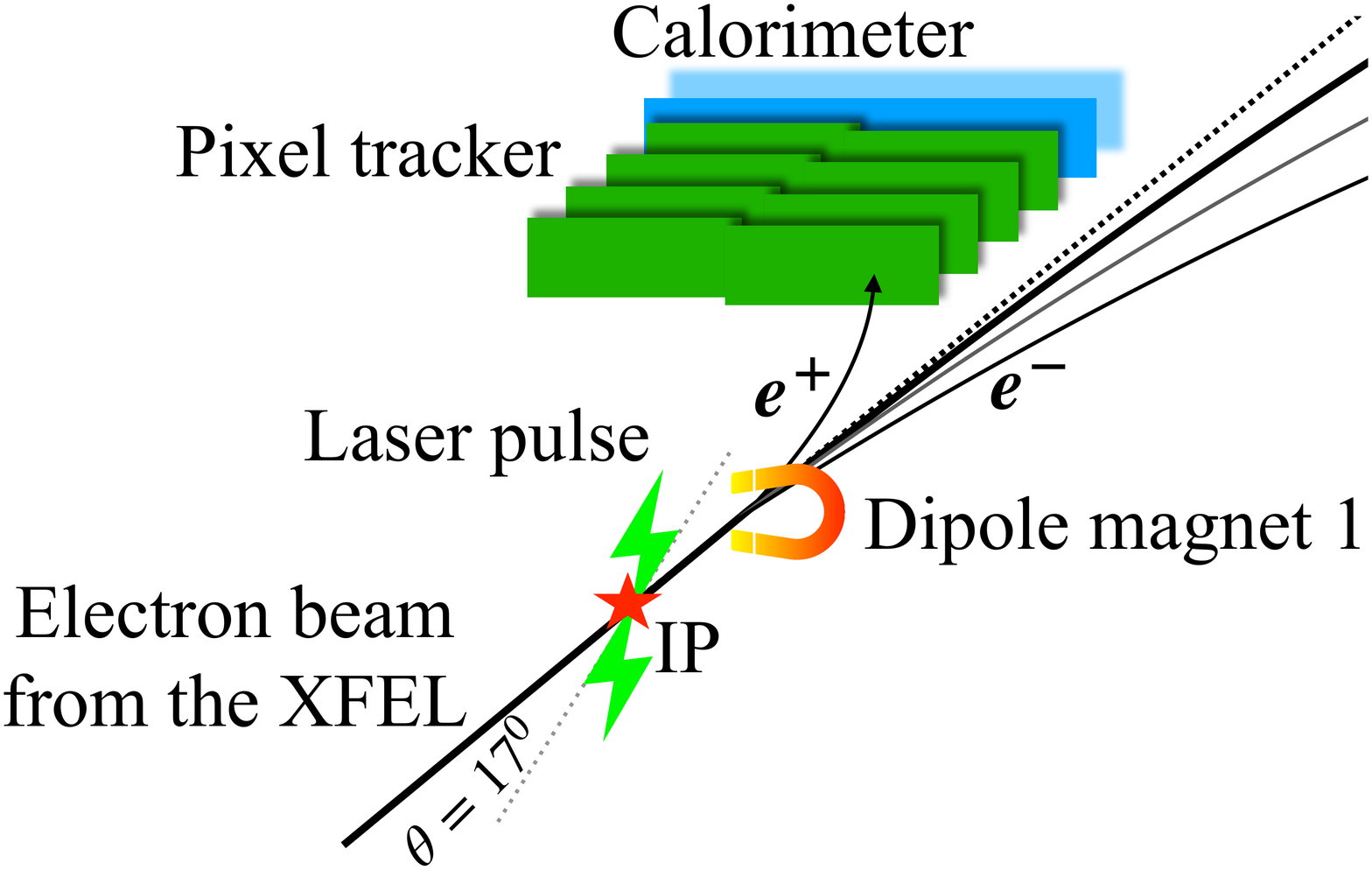}
\caption{Schematic layout of the positron detection system in LUXE for the electron-laser setup. Adapted from Ref.~\cite{Abramowicz:2021zja}. The angle $\theta$ represents the crossing angle of the Eu.XFEL and laser beams.}
\label{fig:LUXEsetup}
\end{figure}

This work focuses on the reconstruction of the electron-laser collisions. In this setup, the electron beam from the Eu.XFEL is guided to the interaction point (IP), where it collides with a laser beam. The experiment plans to start taking data with a 40~TW laser, which will later be upgraded to reach 350~TW.
The electrons and positrons produced in the electron-laser interactions are deflected by a 0.95~T dipole magnet and then detected by a positron detection system, as shown in Figure~\ref{fig:LUXEsetup}.
\footnote{LUXE uses a right-handed coordinate system with its origin at the nominal interaction point and the $z$-axis along the beam line. The $y$-axis points upwards, and the $x$-axis points towards the positron detection system.}

The outgoing positrons are detected using a silicon pixel tracking detector. The tracker consists of four layers, each comprising two $\approx 27$~cm long staves placed next to each other, which overlap partially, as illustrated in the figure. The layers are spaced 10~cm away from each other along the beam axis. The average thickness of the staves is $0.357\%$ of a radiation length. Each stave contains nine sensors, composed of $512\times1024$ pixels of size $27\times29~{\mu \textrm{m}}^2$. 
The pixel sensors have a detection efficiency above 99\%, a noise hit rate much below $10^{-5}$ and a spatial resolution of around $5~\mu$m. 

\section{Simulated data}
\label{sec:datasets}

Monte Carlo simulated event samples are used to perform this study.
The calculation for the electron-laser interaction processes was performed with the \texttt{PTARMIGAN}~\cite{Blackburn:2021rqm} Monte Carlo event generation software.
The electron beam parameters were chosen as follows. The incoming electron energy $\varepsilon_e$ is set to $16.5$\,GeV, the beam spot size to $\sigma_x=\sigma_y=5~\mu$m, $\sigma_z=24~\mu$m, and the normalised emittance to $1.4$~mm$\cdot$mrad. 
The simulation of the laser assumes a 40~TW laser, an energy after compression of 1.2~J and a pulse length of 30~fs. The laser pulse is modelled as having a Gaussian profile both in the longitudinal and in the transverse direction.  The laser spot waist, which for a Gaussian pulse corresponds to $2\sigma$ in intensity, decreases with $\xi$ and varies between 6~$\mu$m and 3~$\mu$m.

The particles produced in the electron-laser interactions are propagated through the dipole magnet and tracking detector using a custom fast simulation that was developed for this study. The fast simulation uses parameterised smearing functions to model the effects of multiple scattering and detector resolution. Furthermore, a simplified detector layout is considered. In this layout, the four detection layers are not split into two overlapping staves, but simply have a double length with no discontinuities.

\begin{figure}[h]
\centering
\includegraphics[width=0.48\textwidth]{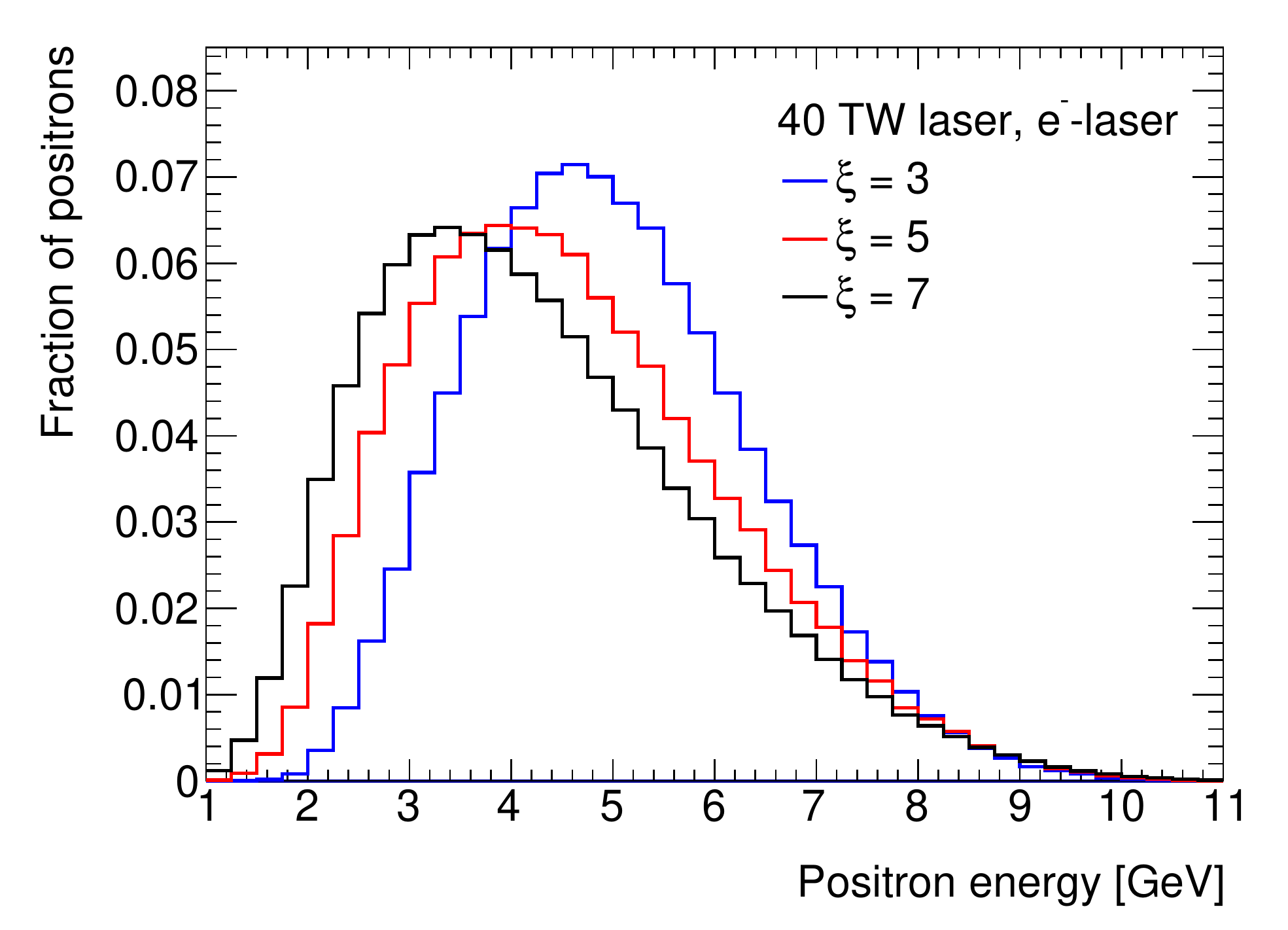}\includegraphics[width=0.5\textwidth]{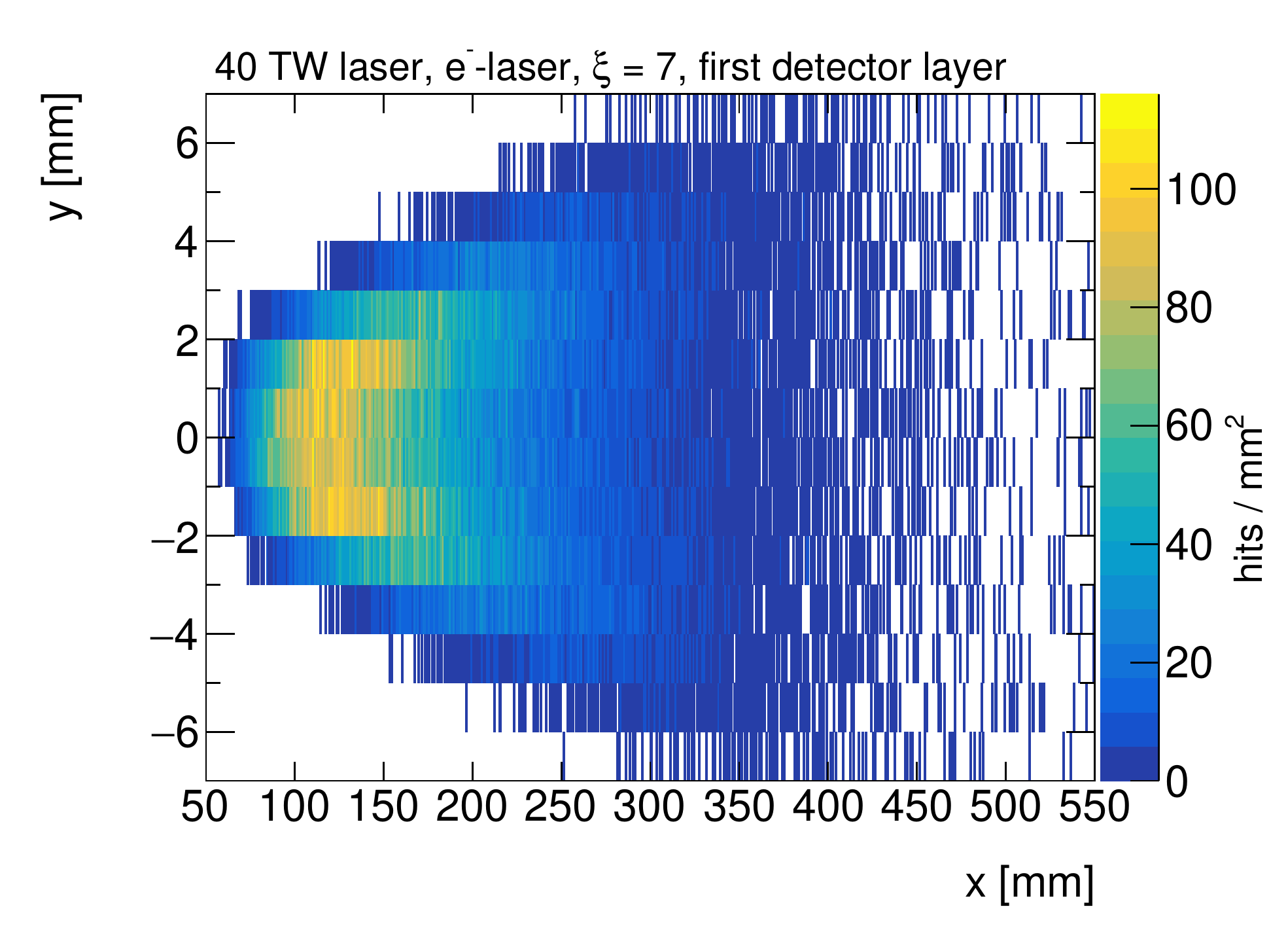}
\caption{Left: Positron energy distribution for different values of $\xi$, normalised to unit area. Based on Ref.~\cite{Abramowicz:2021zja}, using the data sets generated for this work. Right: Number of hits/mm$^2$ in the first detector layer as a function of the $x$ and $y$ coordinates for $\xi=7$.}
\label{fig:simulation}
\end{figure}

To perform these studies, data sets corresponding to electron-laser interactions were generated with $\xi$ values ranging from three to seven and a laser power of 40~TW. This corresponds to positron multiplicities ranging between $1 \times 10^{2}$ and $7 \times 10^{4}$. 
Figure~\ref{fig:simulation} shows the resulting expected positron energy distribution for the three generated $\xi$ values (left) and the number of hits/mm$^2$ in the first detector layer as a function of the $x$ and $y$ coordinates for $\xi=7$ (right). The double-peaked structure visible in the $xy$ plane reflects the initial positron momentum distribution along the $y$-axis at the interaction point.

\section{Methodology}
\label{sec:methods}

The starting point for the pattern recognition are either doublets or triplets, defined as a set of two or three hits in consecutive detector layers. A pre-selection is applied to the initial doublet or triplet candidates to reduce the combinatorial candidates while keeping the efficiency as close as possible to 100\% for the doublets and triplets matching with a real positron. 
Doublets are formed first and are required to satisfy a pre-selection based on the ratio $\delta x/x_0$, where $\delta x$ is the difference of the $x$ coordinates for the two hits composing the doublet, while $x_0$ indicates the $x$ coordinate on the detector layer closest to the interaction point. A window of three standard deviations around the expected mean value of $\delta x/x_0$ for true doublets, as determined in the simulation, is used for this selection. This requirement ensures that the particles come from the IP.
Triplets are subsequently constructed by combining doublet candidates with a requirement on the maximum angle difference $\delta \theta = \sqrt{\delta\theta_{xz}^2 + \delta\theta_{yz}^2}$ of the doublet pairs. The maximum scattering threshold is chosen to be 1~mrad and was optimised taking into account multiple scattering with the detector material. Since triplets consist of three hits, they are formed either from the first to the third layer or from the second to the fourth layer. 

\begin{figure}[h]
\centering
\includegraphics[width=0.5\textwidth]{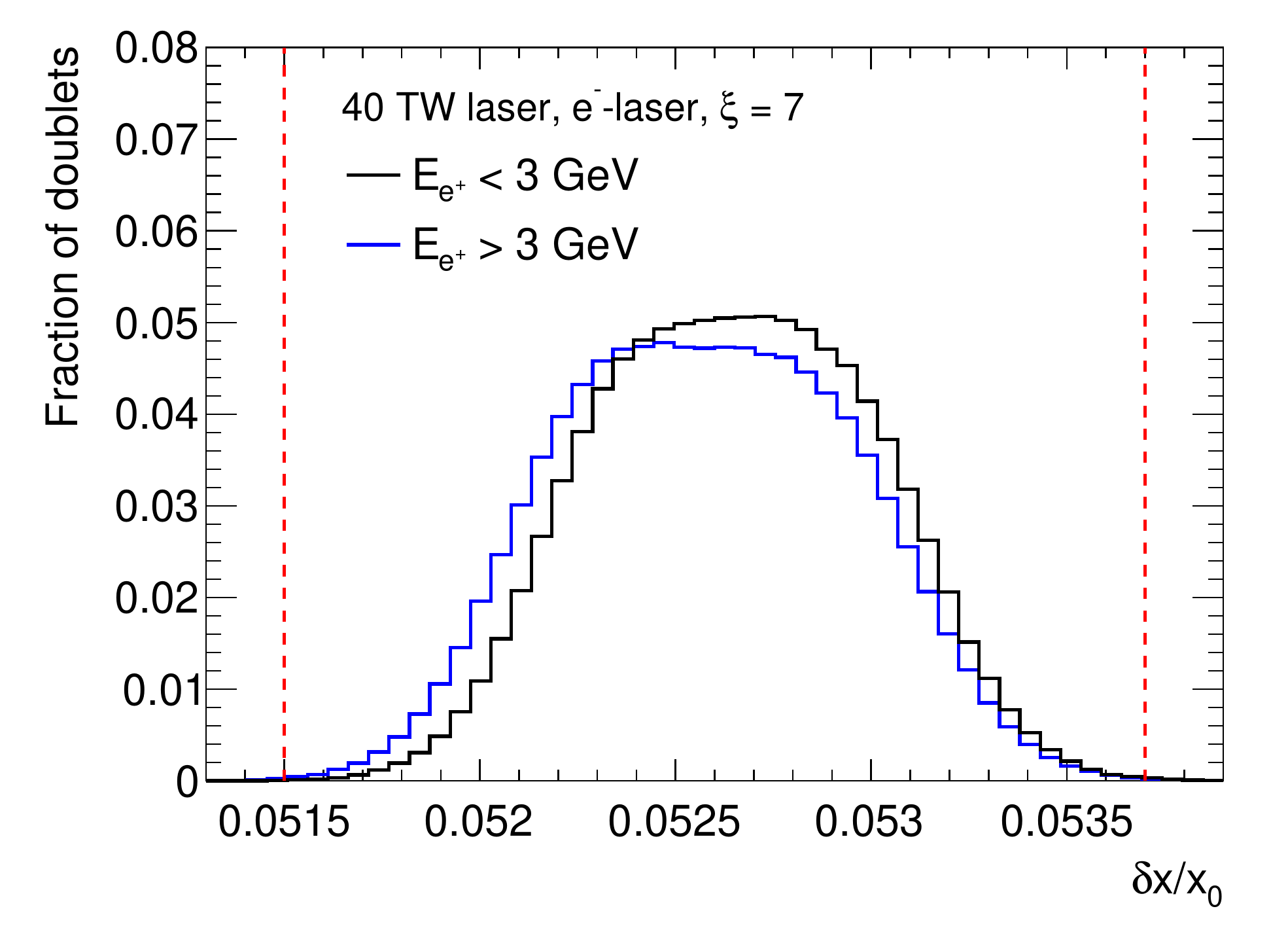}\includegraphics[width=0.5\textwidth]{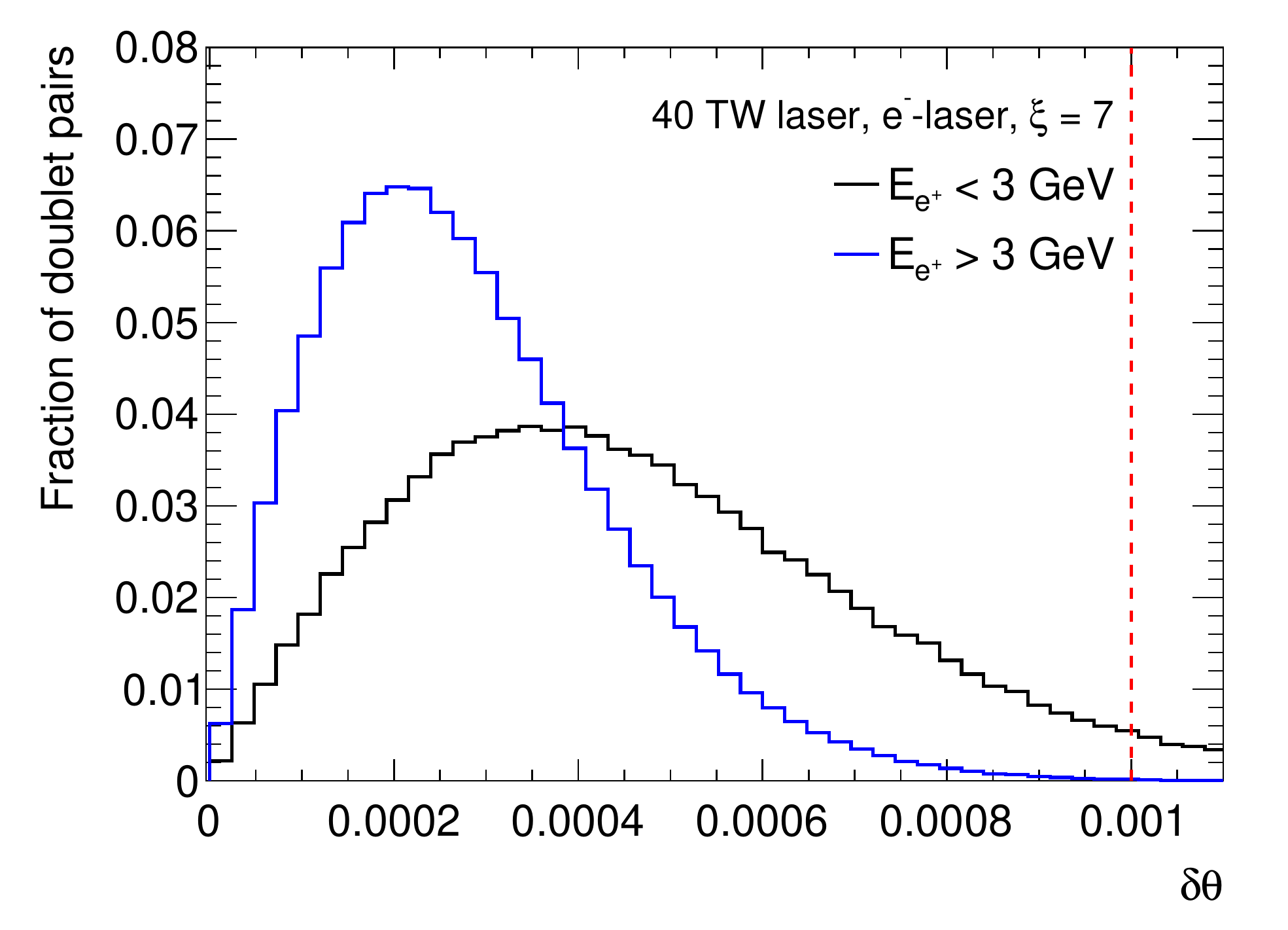}
\caption{Left: Distribution of doublet $\delta x/x_0$ with red dashed lines indicating the range of the pre-selection. Right: Distribution of angle difference $\delta \theta$ for the doublet pairs composing the triplets with a red dashed line indicating the upper limit allowed by the pre-selection.}
\label{fig:preselection}
\end{figure}

\begin{figure}[h]
\centering
\includegraphics[width=0.5\textwidth]{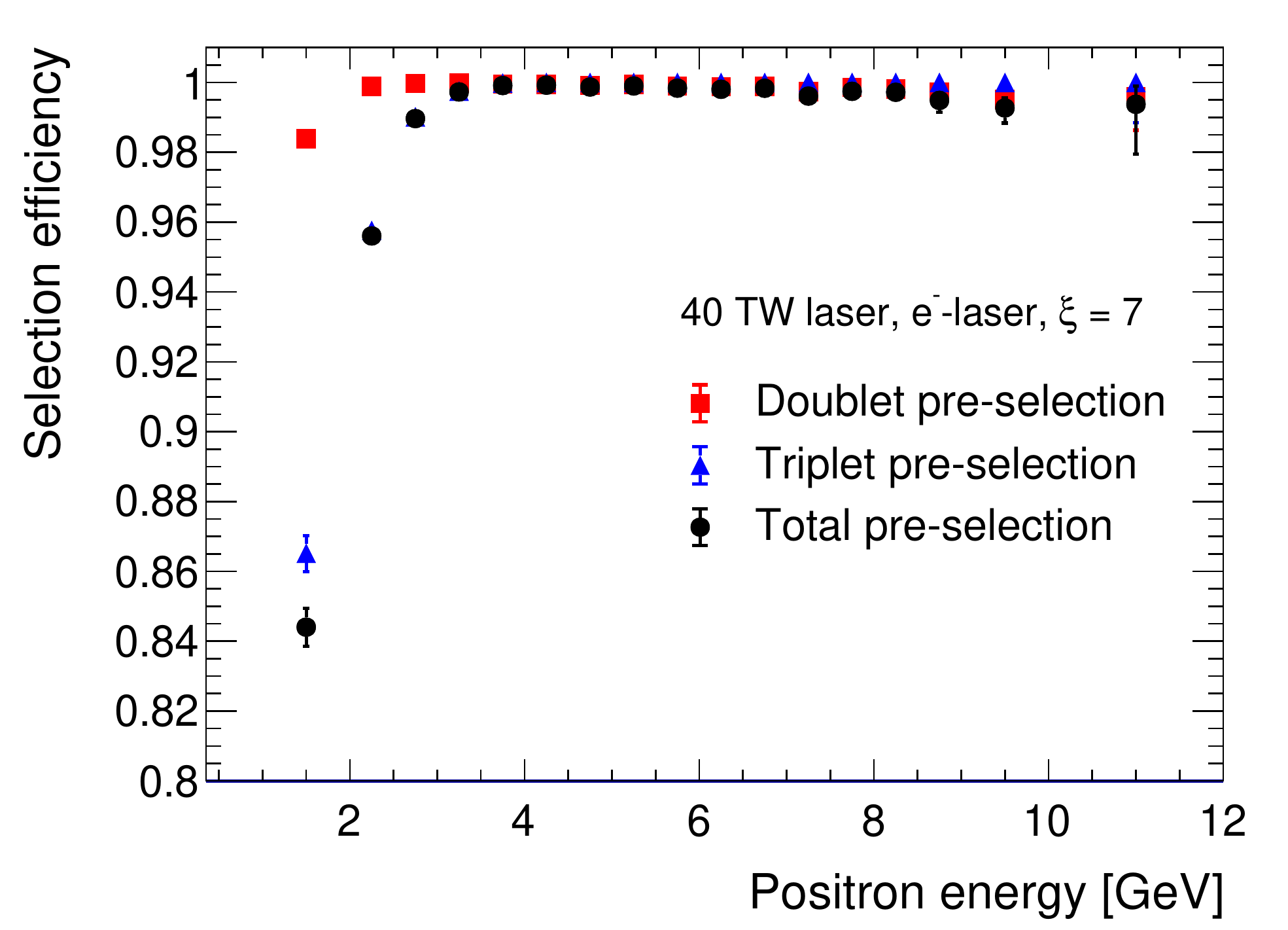}\includegraphics[width=0.5\textwidth]{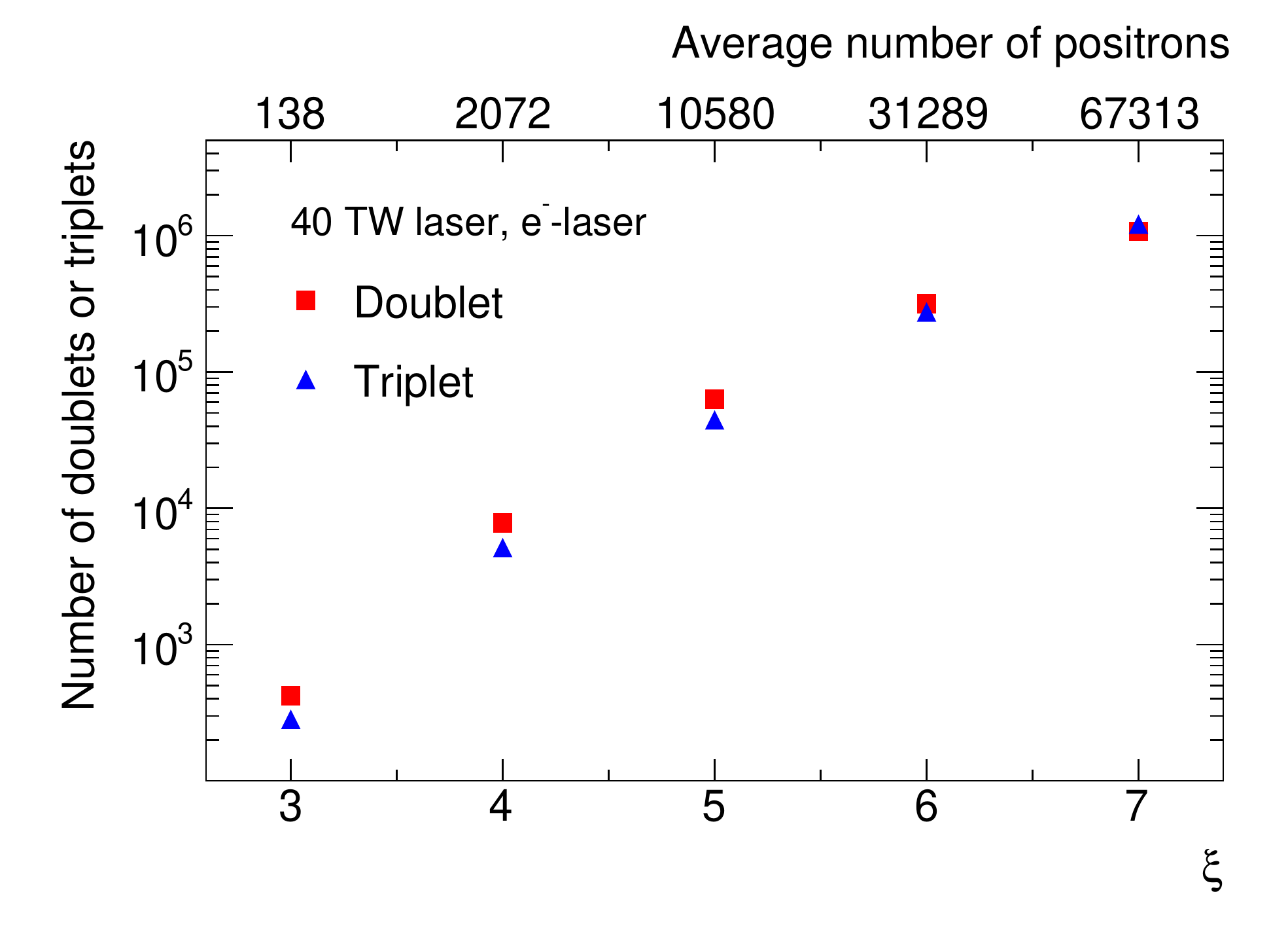}
\caption{Left: Doublet and triplet-finding efficiency as a function of the positron true energy. The combined efficiency is also shown. Right: Doublet and triplet multiplicities as a function of $\xi$ (lower $x$-axis), corresponding to the average number of positrons (upper $x$-axis).}
\label{fig:preselection_eff}
\end{figure}

Figure~\ref{fig:preselection} shows the distributions of $\delta x/x_0$ for doublets (left) and $\delta \theta$ for pairs of doublets (right) originating from true positron tracks, shown separately for low-energy ($E_{e^{+}} < 3$ GeV) and high-energy positrons ($E_{e^{+}} > 3$ GeV), as well as the chosen thresholds. The distributions are obtained using $\xi=7$, but are generally $\xi$-independent. The $\delta x/x_0$ distribution shows a slight dependence on positron energy, while the triplet $\delta \theta$ distribution demonstrates that the scattering is more pronounced for lower energy positrons. The resulting pre-selection efficiencies are shown in Figure~\ref{fig:preselection_eff} (left) for both doublet and triplet finding, in the case of electron-laser interaction for $\xi=7$. The pre-selection requirements are found to be nearly fully efficient for the whole energy range, with a moderate efficiency loss, at the level of 16\% for positron energies below 2~GeV, mostly due to multiple scattering with the detector material. Figure~\ref{fig:preselection_eff} (right) also shows the number of doublets and triplets passing the pre-selection criteria as a function of $\xi$.

Three pattern recognition methods are employed and systematically compared to reconstruct tracks from the detector hits. The first method formulates the tracking problem as a quadratic unconstrained binary optimisation (QUBO), similar to the one used in Ref.~\cite{Bapst:2019llh}, which is then processed with quantum algorithms. 
The second method uses a hybrid quantum-classical graph neural network approach~\cite{Tuysuz:2021oai}, but is limited to specific scenarios compatible with the available devices.
Finally, the results obtained with the quantum approaches are compared to an optimised classical approach based on a Kalman filter~\cite{kalman,Billoir:1983mz}, which is taken to be the reference for the state-of-the-art using no quantum computers.

\subsection{Quadratic unconstrained binary optimisation}
\label{subsec:qubo}

In this approach, the pairs of triplet candidates that can be combined to form tracks are identified by solving a QUBO problem.
The QUBO is expressed via the objective function
\begin{equation}
 O =  \sum_i^N \sum_{j<i} b_{ij} T_i T_j +\sum_{i=1}^{N} a_i T_i ,
\label{eq:QUBO}
\end{equation}
where $T_i$ and $T_j$ are triplets of hits and $a_i$ and $b_{ij}$ are real coefficients. The triplets $T_i$ and $T_j$ assume binary values. The solution of the QUBO determines whether each triplet is considered false and rejected, by being set to zero, or true and selected, by being set to one. The linear term of the QUBO weighs the individual triplets by their quality quantified by the coefficient $a_i$. The $a_i$ coefficient is set to the value of $\delta\theta$ scaled to populate the $[-1; 1]$ range. The quadratic term represents the interactions between triplet pairs, where the coefficient $b_{ij}$ characterises their compatibility. The coefficient $b_{ij}$ is computed from the doublets forming the two considered triplets. It is taken to be the norm of the sum of the standard deviations of the doublet angles in the $xy$ and $yz$ planes, translated and scaled to populate the $[-1; -0.9]$ range. If the two triplets are in conflict, the coefficient $b_{ij}$ is set to one. If the triplets are not connected, it is set to zero. 

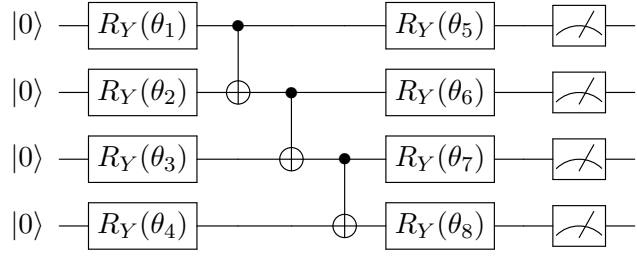
\begin{figure}[h]
\begin{center}
\begin{align*}
  \Qcircuit @C=1em @R=.7em {
  \lstick{\ket{0}} & \gate{R_Y(\theta_1)} & \ctrl{1} & \qw      & \qw      & \gate{R_Y(\theta_5)} & \qw & \meter & \qw \\
  \lstick{\ket{0}} & \gate{R_Y(\theta_2)} & \targ    & \ctrl{1} & \qw      & \gate{R_Y(\theta_6)} & \qw & \meter & \qw \\
  \lstick{\ket{0}} & \gate{R_Y(\theta_3)} & \qw      & \targ    & \ctrl{1} & \gate{R_Y(\theta_7)} & \qw & \meter & \qw \\
  \lstick{\ket{0}} & \gate{R_Y(\theta_4)} & \qw      & \qw      & \targ    & \gate{R_Y(\theta_8)} & \qw & \meter & \qw
  }
\end{align*}
\caption{Layout of the variational quantum circuit using the 
ansatz with $R_Y$ gates and a linear CNOT entangling pattern. For simplicity, only four qubits are shown. \label{fig:circuit}}
\end{center}
\end{figure}

The QUBO in Eq.~\eqref{eq:QUBO} can be mapped to an Ising Hamiltonian by mapping  $T_i \rightarrow (1 + Z_i)/2$, where $Z_i$ is the third Pauli matrix. Minimising the QUBO is equivalent to finding the ground state of the Hamiltonian. 
The Variational Quantum Eigensolver (VQE)~\cite{Peruzzo_2014} method, a hybrid quantum-classical algorithm, was used to find the ground state. In this work, the data is processed using the VQE implementation available in the Qiskit~\cite{matthew_treinish_2023_7757946} library.
Most results rely on classical simulations of quantum circuits, where no sources of noise or decoherence are included, and a simple ansatz with $R_Y$ gates and a linear CNOT entangler is chosen, as shown in Figure~\ref{fig:circuit}. An ansatz with CNOTs between all possible pairs and a single circuit repetition was found to lead to results compatible within statistical uncertainties, but was discarded for simplicity. The selected optimiser is the Nakanishi-Fujii-Todo (NFT)~\cite{Nakanishi_2020} algorithm. The ansatz and optimiser were selected as those leading to the highest track reconstruction efficiency in previous work~\cite{Crippa:2022hbi}.

The number of qubits required to represent the tracking problem as a QUBO is determined by the number of triplet candidates. Due to the limited number of qubits available on the current quantum devices, the QUBO in this work is partitioned into QUBOs of smaller size (referred to as sub-QUBOs) to be solved iteratively. For small enough sub-QUBO sizes, such as the size 7 used in this work, an exact solution using matrix diagonalisation is possible and is used as a benchmark.

\begin{figure}[h]
  \includegraphics[width=38pc]{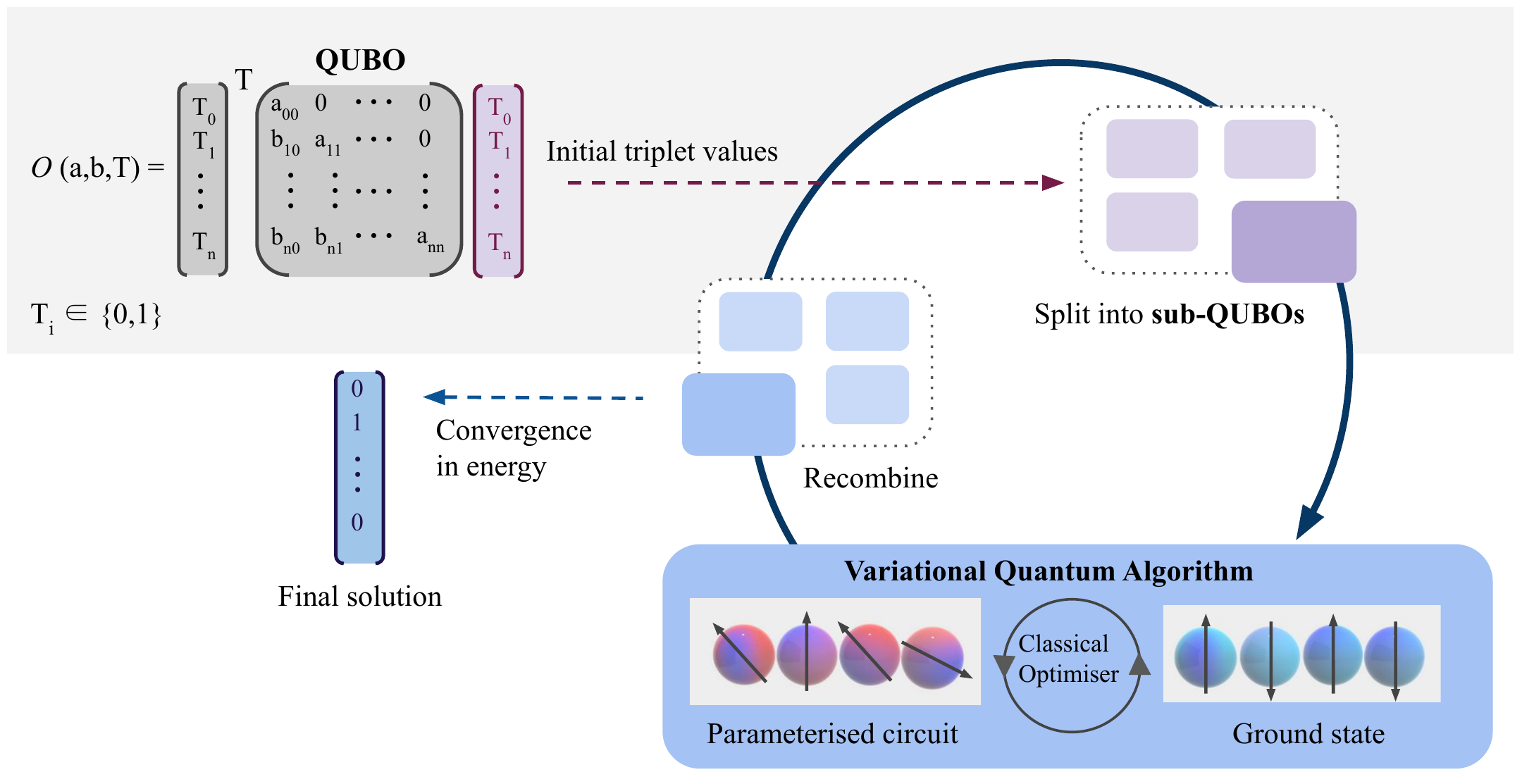}
  \caption{Diagram illustrating the QUBO solving procedure. }
  \label{fig:sketch} 
\end{figure}

Figure~\ref{fig:sketch} summarises the QUBO solving process. At the beginning of the processing, all triplet candidates are set to 1. 
The splitting into sub-QUBOs is done by extracting the sub-QUBO matrices of the desired size, by picking triplets in order of their impact. The impact is defined as the change in the value of the objective function when $T_i \rightarrow 1 - T_i$. 
Each triplet is assigned an additional constant term representing the sum of all interactions with triplets outside of the sub-QUBO to retain sensitivity to the connections outside of each sub-QUBO when computing the value of the objective function. After the sub-QUBOs are solved, the solution is combined. These steps are repeated for a number of iterations. The triplets selected by the QUBO minimisation are retained and matched to form track candidates.

Alternative algorithms for finding the optimal QUBO solution, such as the Quantum Approximate Optimisation Algorithm (QAOA)~\cite{Farhi:2014}, were briefly investigated and were found to lead to significantly worse performance. A dedicated optimisation and characterisation of the results of such alternative algorithms is left to future work.

\subsection{Quantum graph neural network}
\label{subsec:qgnn}
This approach is based on a graph neural network (GNN)~\cite{Farrell:2018cjr,Ju:2021ayy} that consists of both classical neural network layers and quantum circuits.
The graph is constructed from doublets, where the hits are nodes and the connections between the hits are edges. All nodes of consecutive layers are connected and only the ones that satisfy the pre-selection criteria are kept. 
The quantum graph neural network (QGNN) model follows the implementation of Ref.~\cite{Tuysuz:2021oai} and consists of three networks. 
First, the \textit{InputNet} takes the input node features, i.e. the three spatial coordinates, and produces hidden node features. For this purpose, a single fully connected neural network layer that has 10 neurons with a \textit{tanh} activation function is used. 
Second, the \textit{EdgeNet} takes all connected node pairs as input and produces a scalar edge feature for each of them using a \textit{sigmoid} activation function. This will later be the prediction score of the model for each doublet, as this model is essentially a segment classifier. \textit{Circuit 10} with two layers and 10 qubits is selected for this task based on previous work~\cite{Tuysuz:2021oai}. Each layer of this circuit uses $R_{Y}$ gates and linear CNOT entanglers between all possible pairs of qubits. Third, the \textit{NodeNet} considers each node and its connecting nodes to update the hidden node features. The architecture of the \textit{NodeNet} is similar to \textit{EdgeNet}, but it uses the \textit{tanh} activation function for the last layer, as the \textit{NodeNet} is an intermediate step, and \textit{sigmoid} activation functions are known to lead to vanishing gradients. 

The quantum graph neural network (QGNN) model first starts with the \textit{InputNet}. Then, the \textit{EdgeNet} and the \textit{NodeNet} are applied alternately four times to allow the node features to be updated using farther nodes, as determined in a scan of the optimal model parameters. 
At the end, the \textit{EdgeNet} is applied one last time to obtain the predictions for each doublet connection. Finally, the edges are discarded if the prediction value is less than a fixed threshold (chosen to be 0.5 in our simulations) and the rest are retained and used to form track candidates.

\subsection{Combinatorial Kalman filter}
\label{subsec:ckf}

A tracking algorithm based on A Common Tracking Software (ACTS) toolkit~\cite{Ai:2021ghi} with the combinatorial Kalman Filter (CKF) technique for track finding and fitting is used as a benchmark. In this classical tracking method, track finding starts from seeds, which are the triplets formed from the first three detector layers.
To avoid a combinatorial growth in the number of seeds at high particle density, further constraints are placed on seeds sharing the same hits by prioritising the better-aligned seeds. An initial estimate of track parameters is obtained from the seed and is used to predict the next hit and is updated progressively, with the measurement search performed at the same time as the fit. 

\subsection{Final track selection}
\label{sec:trksel}

A final step in the track reconstruction is common to all considered methods. 
Track candidates are required to have four hits and, as explained in the previous subsections, can be found with the QUBO approach that combines triplets into quadruplets (see Section~\ref{subsec:qubo}), by employing the QGNN approach that combines doublets into quadruplets (see Section~\ref{subsec:qgnn}), or by using the classical CKF method (see Section~\ref{subsec:ckf}). After finding these track candidates, the final tracks now have to be selected among these candidates, using a final step explained in the following.

\begin{figure}[h]
\begin{center}
  \includegraphics[width=0.5\textwidth]{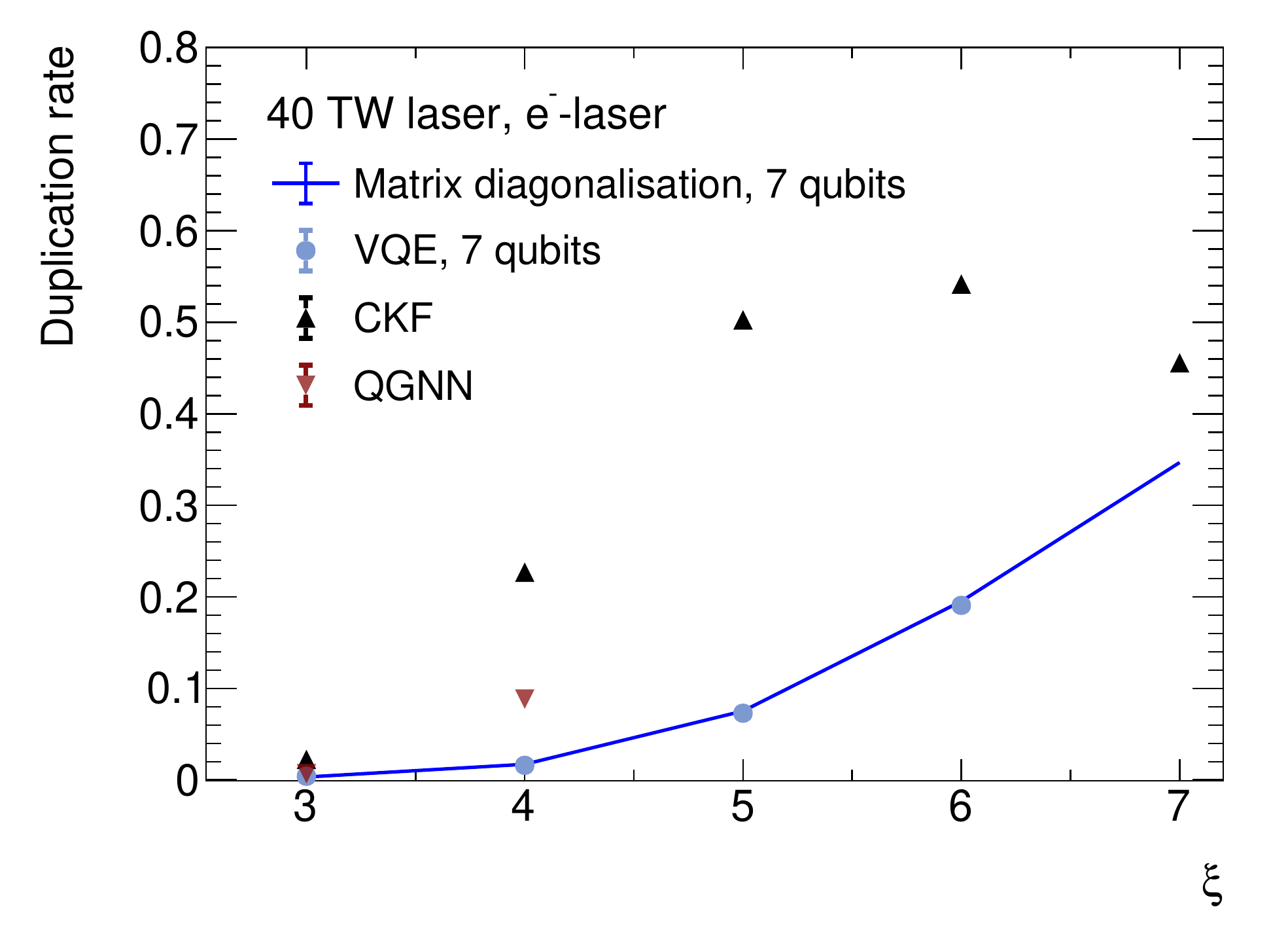}\includegraphics[width=0.5\textwidth]{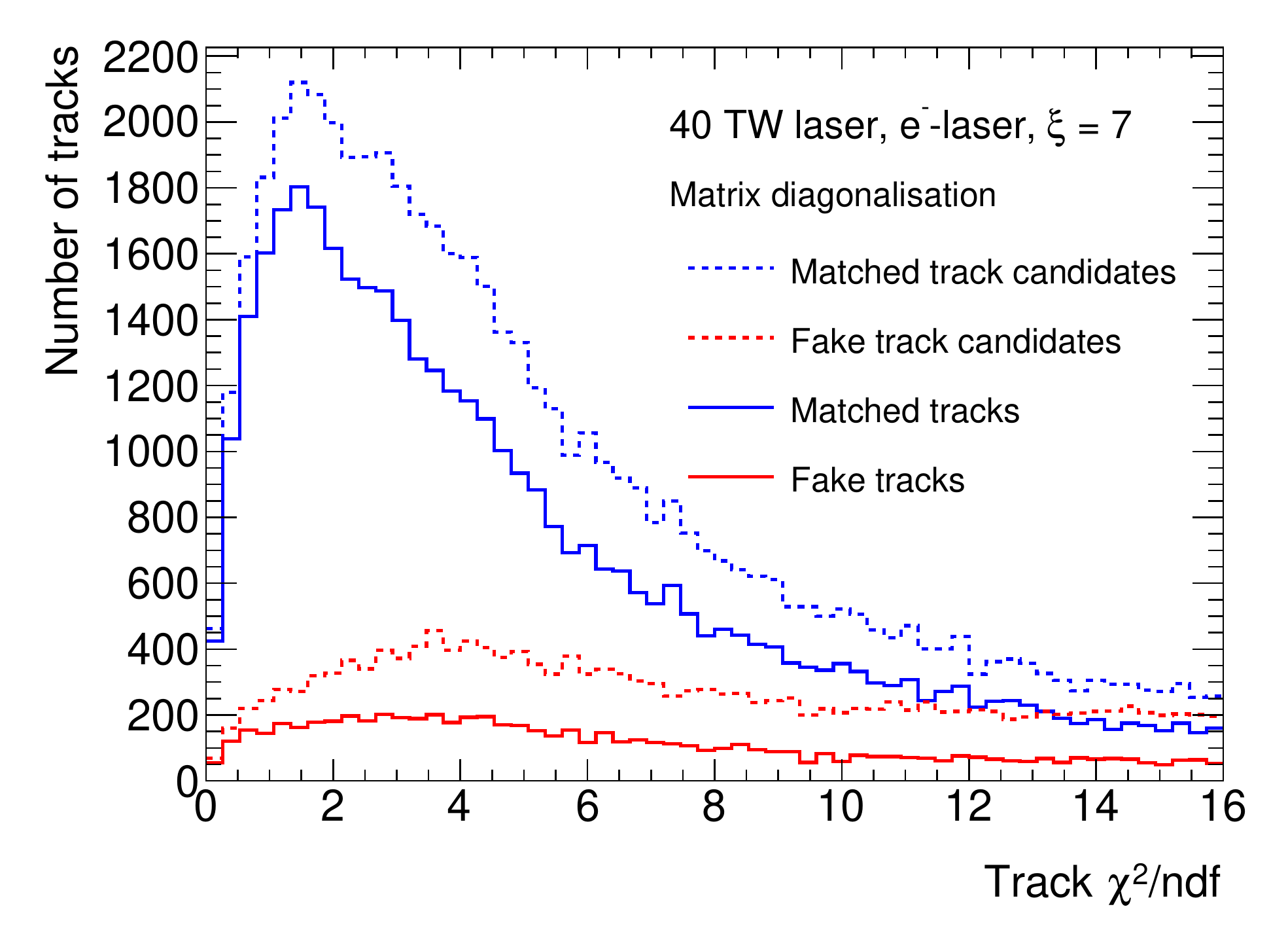}
  \caption{Left: Duplication rate as a function of $\xi$. The results from the exact matrix diagonalisation are shown as a line to help the comparison between the methods. The results based on hybrid quantum-classical methods rely on classical simulations of quantum devices. The decrease in duplication rate for CKF for $\xi>5$ is due to the limit set on seeds with shared hits and an overall decrease in tracking efficiency in this scenario. See Figure~\ref{fig:preselection_eff} (right) for the number of positrons corresponding to each $\xi$. Right: Distribution of the $\chi^2$ divided by the number of degrees of freedom for fitted track candidates found using the QUBO approach with exact matrix diagonalisation, shown separately for matched (blue) and fake (red) track candidates. The dashed lines represent the track candidates from the QUBO solution, while the solid lines represent the selected tracks after the resolution of reconstruction ambiguities.}
  \label{fig:overlap_and_chi2}
  \end{center}
\end{figure}

The track candidates are fitted to straight lines with the least-square method, as the particles propagate through the tracking detector in absence of a magnetic field. A track candidate is considered matched if it has at least three out of four hits matched to the same particle. Figure \ref{fig:overlap_and_chi2} (left) shows the duplication rate, i.e. the fraction of matched particles that are matched to more than one track candidate, as a function of $\xi$. The substantially larger duplication rate of the CKF technique is due to this method being a local approach with no knowledge of the overall BX, unlike the QUBO and QGNN-based approaches.

To resolve the overlaps between the track candidates and to reject fake tracks, an ambiguity resolution step is performed. The track candidates are scored based on the $\chi^2$/ndf of the track fit and the number of shared hits with other track candidates. The track candidates with the most shared hits are evaluated first. They are compared to the other track candidates sharing the same hits, and the ones with worse $\chi^2$/ndf of the track fit are rejected. The procedure is repeated until all remaining tracks have up to one shared hit. 
Figure~\ref{fig:overlap_and_chi2} (right) shows the effect of the ambiguity resolution on matched and fake tracks for a QUBO solved using matrix diagonalisation in a BX with $\xi=7$. This scenario was selected to show the effect for the highest particle multiplicity considered in this work.

\section{Results}
\label{sec:results}

\subsection{Studies with classical hardware}

The results presented in the following are obtained on classical hardware, including classical simulations of quantum hardware. A set of studies performed on quantum hardware (ibm\_nairobi) will be presented in Section~\ref{sec:realquantum}.
The performance of various tracking methods is assessed using the efficiency and the fake rate as metrics, which are computed on the final set of tracks.
The efficiency and fake rate are defined as
\begin{equation}
  \textrm{Efficiency} = \frac{N_{\rm tracks}^{\rm matched}}{N_{\rm tracks}^{\rm generated}} \qquad \text{and}\qquad
  \textrm{Fake rate} = \frac{N_{\rm tracks}^{\rm fake}}{N_{\rm tracks}^{\rm reconstructed}}\, .
\end{equation}

\begin{figure}[h]
\includegraphics[width=0.5\textwidth]{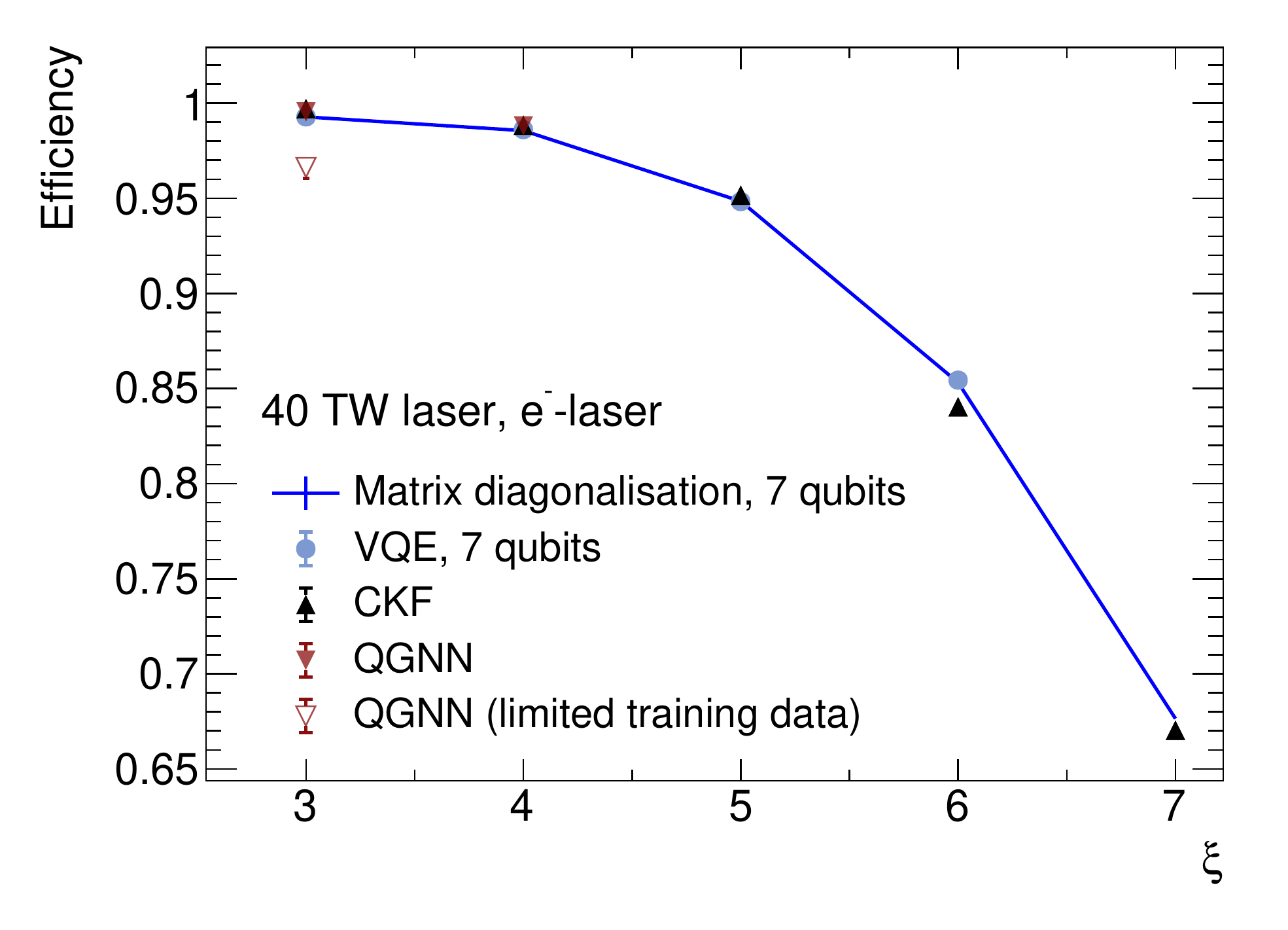}\includegraphics[width=0.5\textwidth]{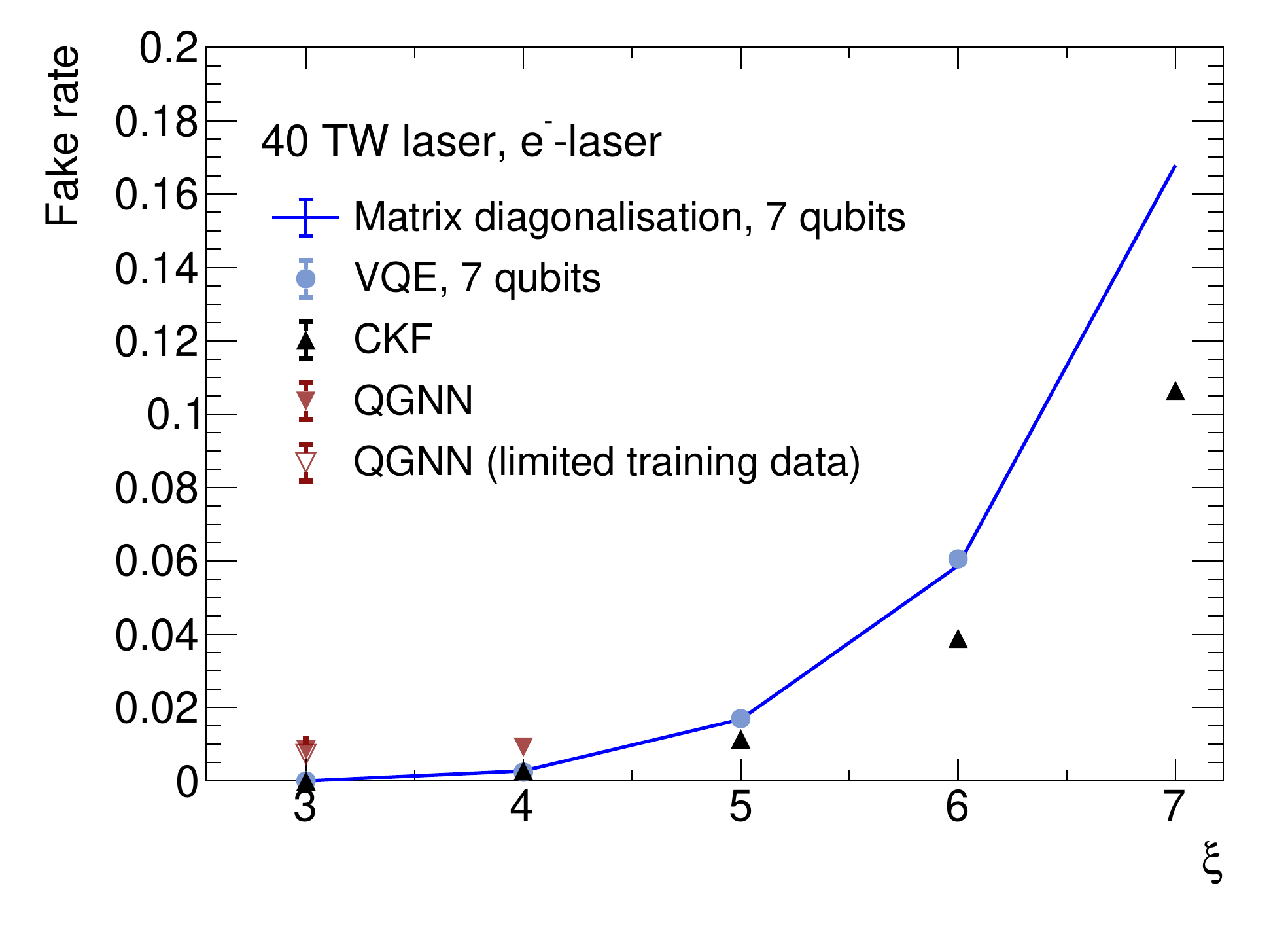} 
\caption{Left: Track reconstruction efficiency as a function of the field intensity parameter $\xi$. Right: Track fake rate as a function of $\xi$. The results from the exact matrix diagonalisation are shown as a line to help the comparison between the methods. The empty triangles show the results of a QGNN training limited to 90 BXs. See Figure~\ref{fig:preselection_eff} (right) for the number of positrons corresponding to each $\xi$. The results based on hybrid quantum-classical methods rely on classical simulations of quantum devices.}
\label{fig:effvsxi_track} 
\end{figure}

Figure~\ref{fig:effvsxi_track} shows the average track reconstruction efficiency (left) and fake rate (right) as a function of the laser field intensity parameter $\xi$ for all tested approaches: QUBO-based tracking (both with the approximate solution obtained with VQE and the exact solution via matrix diagonalisation), QGNN-based tracking, and conventional CKF-based tracking.

The performance of CKF-based tracking is used as a state-of-the-art benchmark. The excellent performance of the classical method deteriorates with $\xi$, because of the increasing hit density. The results using the exact matrix diagonalisation to solve the QUBO are well aligned with the CKF algorithm and achieve a higher efficiency by 1--2\% for large values of $\xi$ at the cost of an increase in the fake rate of approximately a factor of two. The rate of purely combinatorial tracks, i.e. tracks reconstructed from four hits belonging to four distinct positrons, accounts for about 50\% of the total fake rate, independently of the reconstruction algorithm considered.
The results for VQE are in excellent agreement, within the statistical uncertainties, with those from the matrix diagonalisation. 

The results for the QGNN-based tracking are shown up to $\xi=4$, above which simulating the quantum circuits becomes computationally prohibitive with the currently available resources. The reconstruction efficiency is found to be compatible with the other methods, with a substantially higher fake rate. Further work aimed at optimising the selection of the \textit{EdgeNet} predictions could mitigate this effect. 
The QGNN results were validated by implementing a classical GNN~\cite{Farrell:2018cjr,Ju:2021ayy} with the same architecture, but with 128 node hidden features, finding excellent agreement. For $\xi=3$, two values of QGNN efficiency are shown. The empty triangle is the result based on 100 BXs, i.e. the same number of BX used to evaluate the performance of the CKF and QUBO-based methods, using 90\% of the data for the training of the model and 10\% for the inference. Because of the modest particle multiplicity expected at $\xi=3$, the number of true tracks used in the QGNN training is too small to obtain an optimal result. The full triangles show the efficiency obtained with the QGNN training based on data generated with $\xi=4$, which corresponds to a substantially larger set of true tracks, restoring a higher efficiency. 

\begin{figure}[h]
\centering
\includegraphics[width=0.5\textwidth]{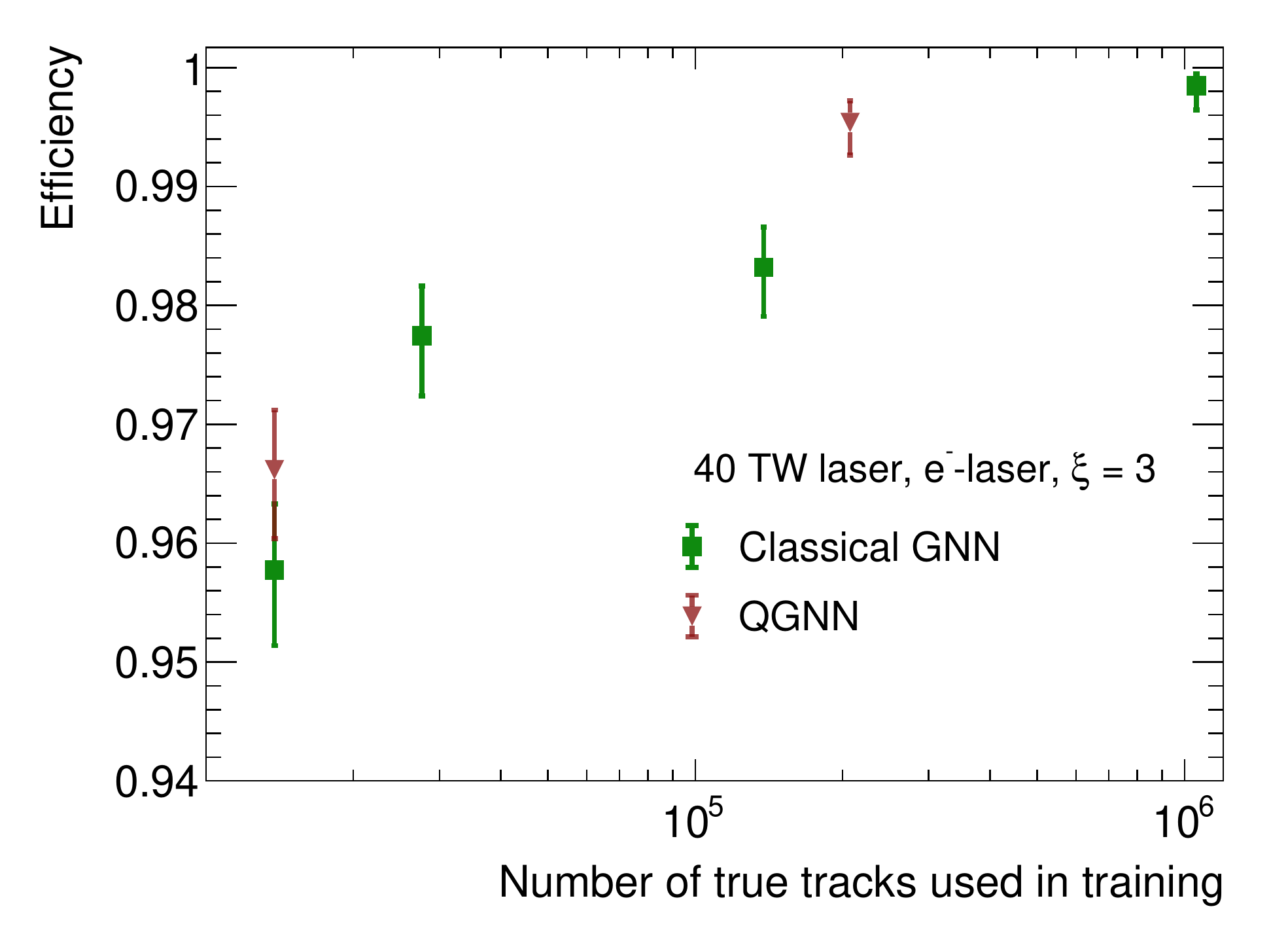}
\caption{Track reconstruction efficiency as a function of the number of true tracks used in the training of a classical GNN (green squares) and the QGNN (brown triangles) for $e^{-}$-laser collisions with $\xi=3$. The QGNN results rely on classical simulations of quantum hardware.}
\label{fig:effvstraining} 
\end{figure}

The dependency of the track reconstruction efficiency on the GNN-based approaches was further studied in $e^{-}$-laser collisions with $\xi=3$, comparing the results obtained with the QGNN and with a classical GNN for different numbers of true tracks used in the training. The findings are presented in Figure~\ref{fig:effvstraining}. The efficiency results for the largest track multiplicity of both GNNs are obtained performing the training on events with a larger value of $\xi=5$ for the classical GNN and $\xi=4$ for the QGNN. All other data points are obtained by increasing the number of BXs considered at $\xi=3$. While it is not expected for the QGNN and the classical GNN to perfectly overlap in performance because of the slightly different model architectures, the results show compatible trends when considering additional data for a fixed value of $\xi$ and using models trained on BXs with larger $\xi$. 

\begin{figure}[h]
\includegraphics[width=0.5\textwidth]{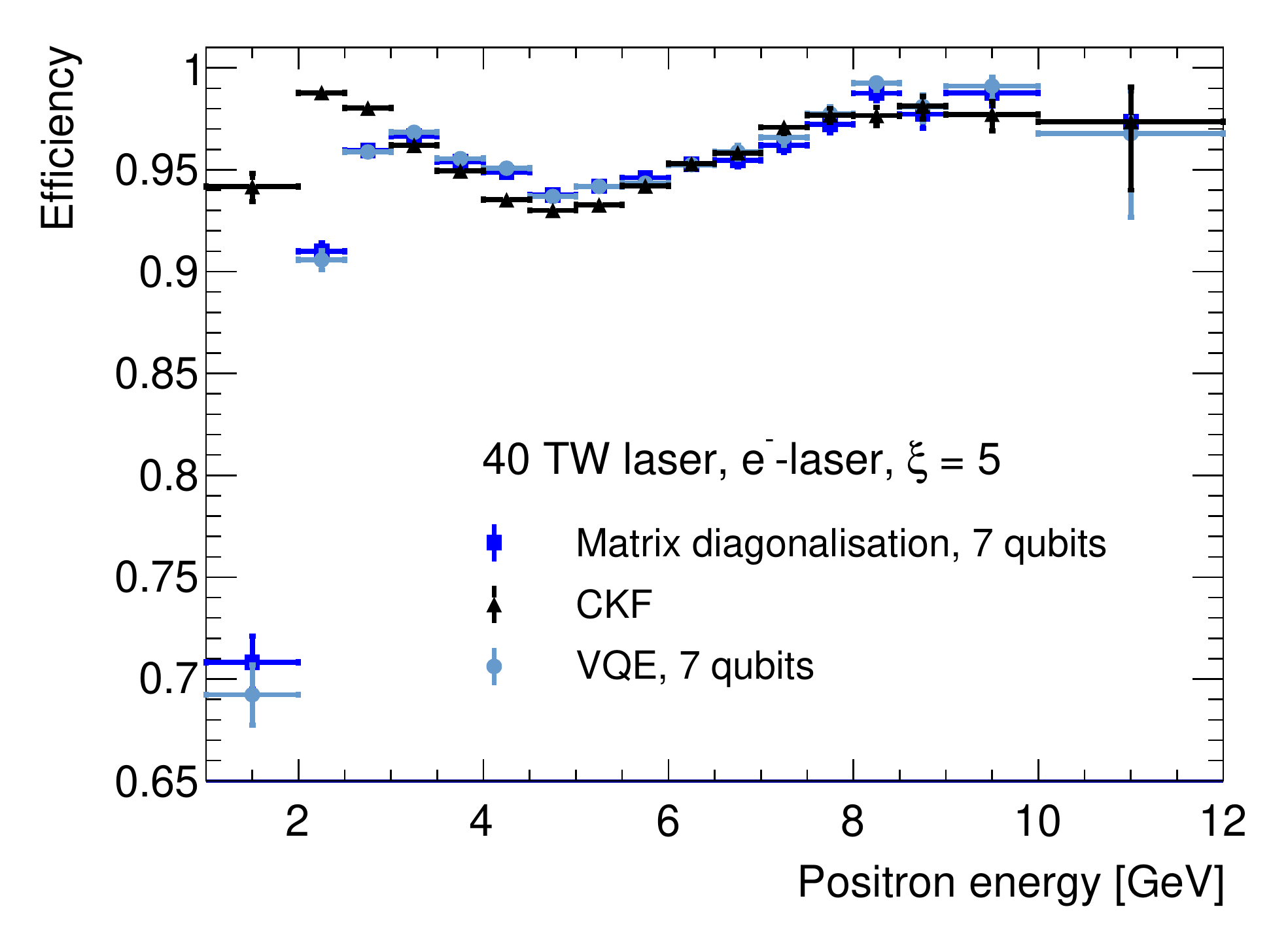}
\includegraphics[width=0.5\textwidth]{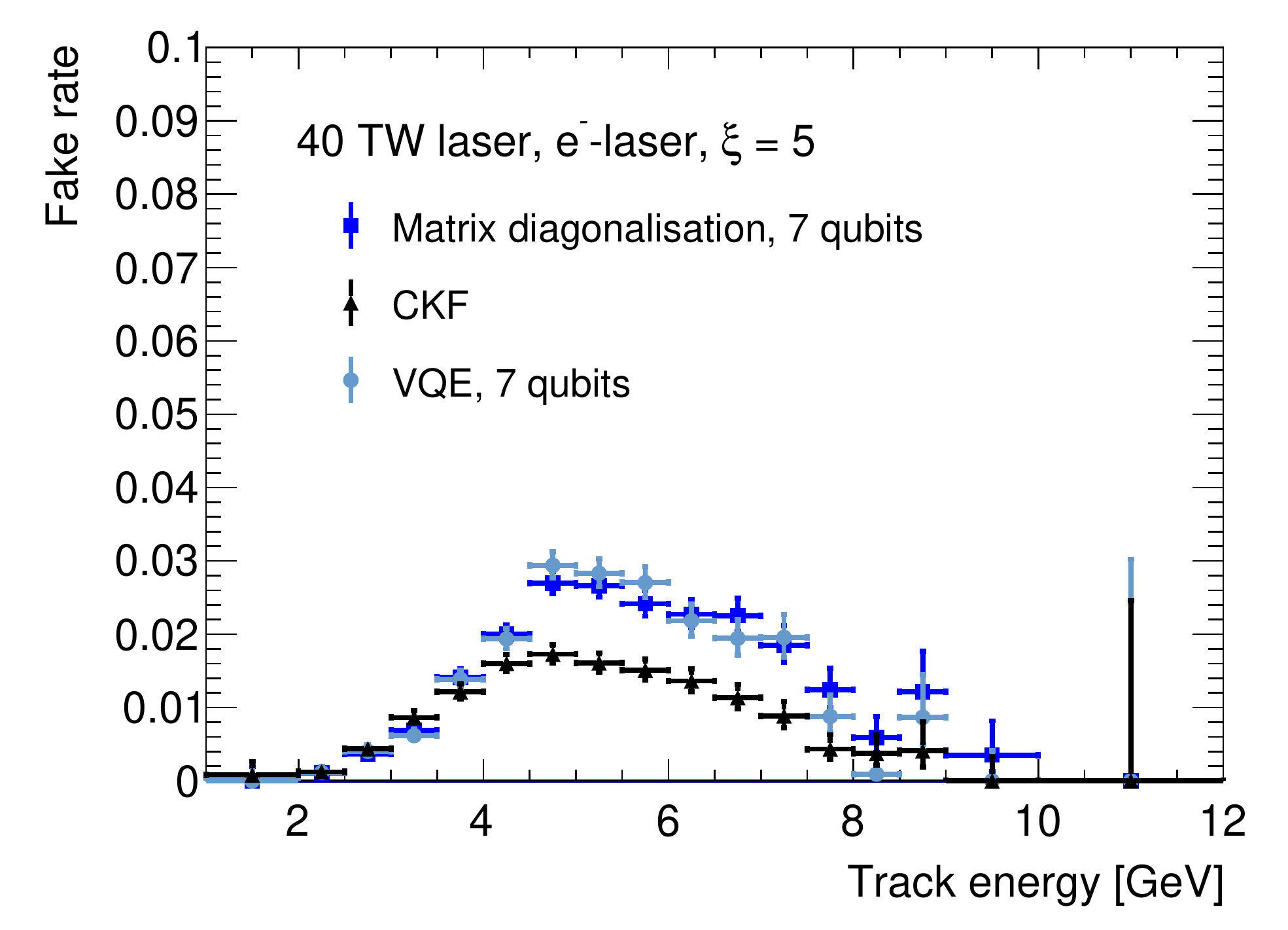}
\caption{Left: Track reconstruction efficiency as a function of the positron true energy for $\xi=5$. Right: Track fake rate as a function of the measured track energy for $\xi=5$. The results based on hybrid quantum-classical methods rely on classical simulations of quantum devices. On average, about 10500 positrons are expected to be produced in a BX with $\xi=5$.}
\label{fig:effvse_track} 
\end{figure}

Figure~\ref{fig:effvse_track} shows the track reconstruction efficiency (left) and fake rate (right) as a function of the true positron energy for the case of $\xi=5$, for the CKF and QUBO-based methods.
The methods show similar behaviours, with a decrease in the region corresponding the highest detector occupancy. Because of effects coming from the propagation through the magnetic field and from the longitudinal size of the interaction region, the maximum occupancy shown in Figure~\ref{fig:simulation}, does not correspond to the maximum of the positron energy distribution. 
The reduced efficiency of the QUBO-based methods for positrons with an energy below 3~GeV is dominated by the pre-selection efficiency shown in Figure~\ref{fig:preselection_eff} (left).

The average energy resolution of the reconstructed tracks was also compared between the different methods. The track energy resolution was found to be 0.5\% and independent of the reconstruction method within the statistical uncertainty of the analysed data set. 

\FloatBarrier
\subsection{Studies with quantum hardware}
\label{sec:realquantum}

This section presents a detailed assessment of the performance of the VQE algorithm on QUBOs of size seven, chosen to be the same as the sub-QUBO size used for the results based on classically simulated VQE in Section~\ref{sec:results}.

A QUBO representing two nearby particles, leading to a total of seven triplets, was selected for this test. The VQE method was applied first in an exact classical simulation assuming an ideal quantum device with shot noise only, then in a classical simulation involving a noise model extracted from a snapshot of the measured noise of the ibm\_nairobi device (fake\_nairobi) and finally using real quantum hardware (ibm\_nairobi). 

For each of these scenarios, 512 circuit evaluations (shots) were considered. When performing the computations with fake\_nairobi and ibm\_nairobi, a measurement error mitigation based on the generation of a calibration matrix was used~\cite{Bravyi:2021,Acharya:3400302.3415684}. The readout error probabilities were calibrated every 30 function evaluations of the optimiser. 

\begin{figure}[htb]
\centering
\includegraphics[width=0.6\textwidth]{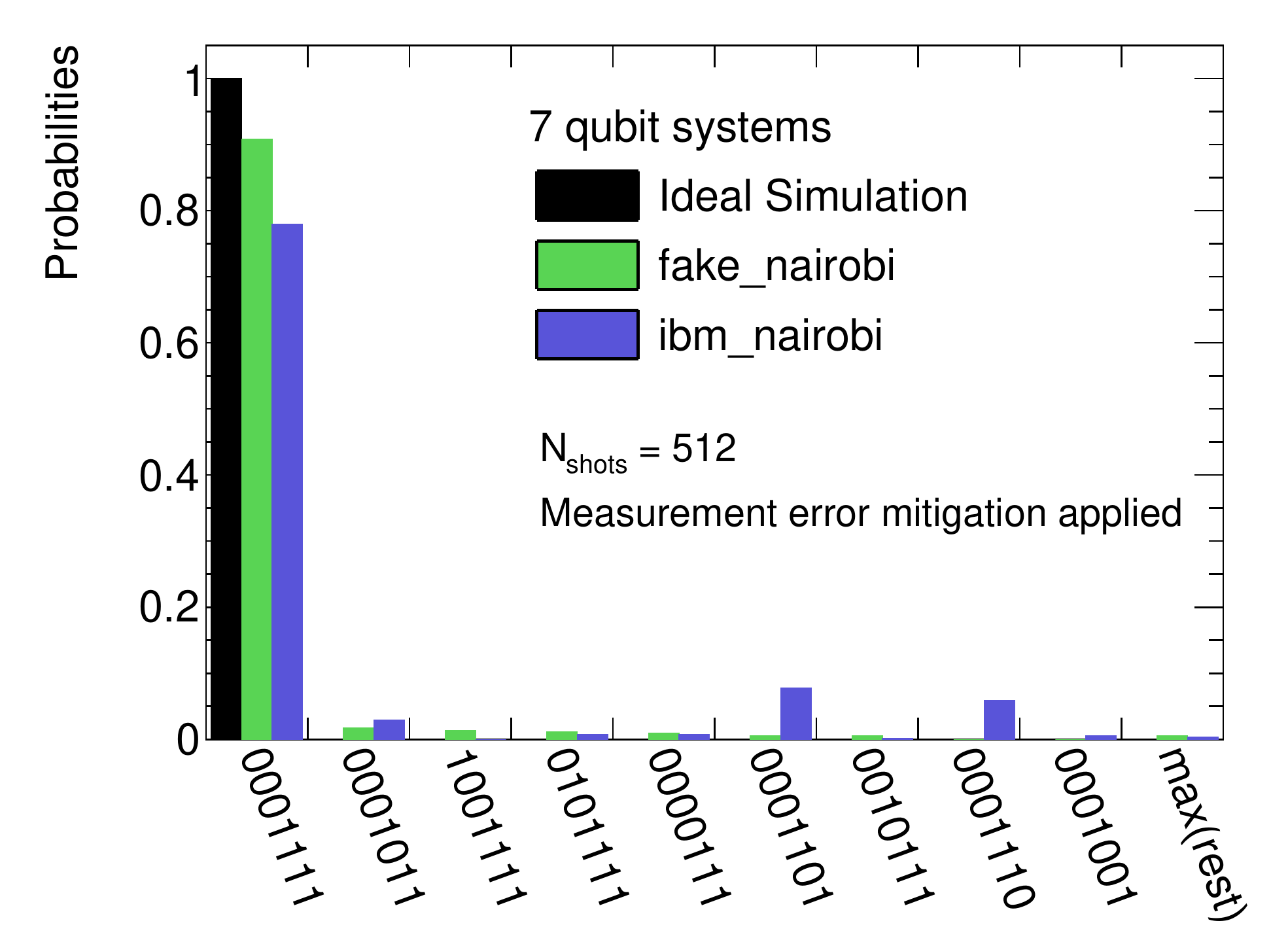}
\caption{Distribution of the VQE results for a test QUBO composed of seven triplets. The blue bars indicate the results obtained from 512 shots on the ibm\_nairobi quantum computer, compared with a realistic (green bars) and an ideal (blue bars) classical simulation of the same system. The results from the realistic classical simulation and from ibm\_nairobi use a measurement error mitigation technique based on the generation of a calibration matrix~\cite{Bravyi:2021,Acharya:3400302.3415684}.}
\label{fig:realdevice}
\end{figure}

Figure~\ref{fig:realdevice} shows the probabilities of the returned results for these three scenarios, where the correct binary solution 0001111 is also the most probable.

\section{Conclusion}
\label{sec:conclusions}

This work investigated the use of hybrid quantum-classical algorithms for particle track reconstruction. Focusing on a VQE approach for a QUBO formulation of track reconstruction and a QGNN approach, the performance of these hybrid quantum-classical methods was compared to results obtained from a state-of-the-art classical tracking method.

In order to produce these results, a standalone fast simulation of the LUXE tracking detector was put in place as well as a software framework to reconstruct tracks up to the maximum number of positrons expected during the data taking with a laser power of 40~TW.

The results were analysed in terms of reconstruction efficiency, fake rate and energy resolution. Hybrid quantum-classical algorithms were found to lead to competitive results when compared to classical algorithms. For large particle multiplicities, a QUBO approach based on VQE using a classical simulation of a quantum device was found to have moderately higher efficiency than classical tracking, but with a significant increase in the fake rate. It was not possible, due to limitations in the computing resources, to evaluate the performance of the approach based on QGNNs beyond a few thousand charged particles.

\section{Outlook}
\label{sec:outlook}

In this work, it was observed that the impact-based processing order leads to a significant fraction of trivially-solvable sub-QUBOs with no interacting triplets. Future work will be aimed at developing alternative algorithms for the sorting of the binary vector representing the triplet candidates and for the splitting of the problem into sub-QUBOs. To further reduce the computation time and the rate of fake tracks reconstructed with this method, future work will focus on optimising the scaling ranges for the $a_i$ and $b_{ij}$ coefficients.

While the initial study of the VQE performance on real quantum hardware (ibm\_nairobi) yielded promising results, a more systematic study of hybrid quantum-classical algorithms using NISQ-era devices will be performed in future work.

Finally, the choice of the optimiser used for VQE has a significant impact on the probability to find the true minimum of the cost function, and a careful optimisation will be required when considering larger sub-QUBO sizes.

\ack{The authors thank the LUXE collaboration for fostering this work. The work by B.H., A.K., F.M., D.S.\ and Y.Y.\ were in part funded by the Helmholtz Association - ``Innopool Project LUXE-QED''. A.C., K.J.\ and C.T.\ are supported in part by the Helmholtz Association - ``Innopool Project Variational Quantum Computer Simulations (VQCS)''. L.F.\ is partially supported by the U.S.\ Department of Energy, Office of Science, National Quantum Information Science Research Centers, Co-design Center for Quantum Advantage (C$^2$QA) under contract number DE-SC0012704, by the DOE QuantiSED Consortium under subcontract number 675352, by the National Science Foundation under Cooperative Agreement PHY-2019786 (The NSF AI Institute for Artificial Intelligence and Fundamental Interactions, http://iaifi.org/), and by the U.S.\ Department of Energy, Office of Science, Office of Nuclear Physics under grant contract numbers DE-SC0011090 and DE-SC0021006. S.K.\ acknowledges financial support from the Cyprus Research and Innovation Foundation under project ``Future-proofing Scientific Applications for the Supercomputers of Tomorrow (FAST)'', contract no.\ COMPLEMENTARY/0916/0048. This work is supported with funds from the Ministry of Science, Research and Culture of the State of Brandenburg within the Centre for Quantum Technologies and Applications (CQTA). This work is funded within the framework of QUEST by the European Union’s Horizon Europe Framework Programme (HORIZON) under the ERA Chair scheme with grant agreement No.\ 101087126. This work has benefited from computing services provided by the German National Analysis Facility (NAF).
\begin{figure}[htb]
    \centering
    \includegraphics[width=0.1\textwidth]{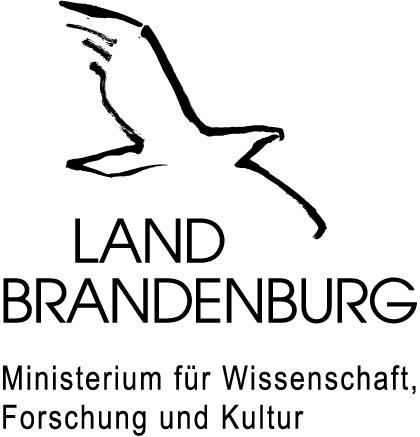}
\end{figure}
}

\section*{References}
\bibliographystyle{report}
\bibliography{references.bib}

\end{document}